\documentclass[epj]{svjour} 

\usepackage{graphicx}
\usepackage{dcolumn}
\usepackage{bm}
\usepackage{verbatim}

\begin{document}

\title{First detection and energy measurement of recoil ions following beta decay in a Penning trap with the WITCH experiment}

\author{M. Beck\inst{1} \and S. Coeck\inst{2} \and V.Yu. Kozlov\inst{2}\thanks{Present address: Karlsruhe Institute of Technology, Institut f\"ur Kernphysik, Postfach 3640, 76021 Karlsruhe, Germany} \and M. Breitenfeld\inst{2} \and P. Delahaye\inst{3} \and P. Friedag\inst{1} \and M. Herbane\inst{2} \and A. Herlert\inst{3} \and I.S. Kraev\inst{2} \and J. Mader\inst{1} \and M. Tandecki\inst{2} \and S. Van Gorp\inst{2} \and F. Wauters\inst{2} \and Ch. Weinheimer\inst{1} \and F. Wenander\inst{3} \and N. Severijns\inst{2}}

\institute{{Westf\"{a}lische Wilhelms-Universit\"{a}t M\"{u}nster, Institut f\"{u}r Kernphysik, Wilhelm-Klemm Str. 9, D-48149, M\"unster, Germany} \and {K.U.Leuven, Instituut voor Kern- en Stralingsfysica, Celestijnenlaan 200 D, B-3001 Leuven, Belgium} \and {Physics Department, CERN, 1211 Geneva 23, Switzerland}}

\titlerunning{Recoil ion measurement after $\beta$ decay}
\authorrunning{M. Beck et al.}

\date{\today}

\mail{marcusb@uni-muenster.de,\\ nathal.severijns@fys.kuleuven.be}

\abstract{
The WITCH experiment (Weak Interaction Trap for CHarged particles) will search for exotic interactions by investigating the $\beta$-$\nu$ angular correlation via the measurement of the recoil energy spectrum after $\beta$ decay. As a first step the recoil ions from the $\beta^-$ decay of $^{124}\mathrm{In}$ stored in a Penning trap have been detected. The evidence for the detection of recoil ions is shown and the properties of the ion cloud that forms the radioactive source for the experiment in the Penning trap are presented.
}

\maketitle

\section{Introduction}

In the Standard Model description of the weak interaction only two
out of five theoretically possible terms are included in the $V-A$
Hamiltonian \cite{Sev06}. Although this still provides a
good description of the experimental data to date, exotic scalar or tensor
type currents have not fully been excluded \cite{Sev06}. To
determine the contributions of the various possible terms in the
Hamiltonian, correlation coefficients in $\beta$ decay are often
measured \cite{Jac57,Car91,Qui93,Ska94,Sev00,Abe08,Har09,Pit09,Wau09}. The $\beta$-$\nu$ angular correlation
coefficient, $a$, has been addressed several times in the past because
of its high sensitivity to these exotic weak currents
\cite{Beh09,Joh63,Sci04,Gor05,Ade99,Bec03,Rod06,Glu05,Vet08,Fle08,Bas08,Poc09}. Since measuring the correlation between the directions of emission of the
$\beta$ particle and the neutrino by directly observing the
neutrino is impossible, $a$ is commonly inferred from a measurement in which the
recoiling nucleus is observed.
The WITCH experiment \cite{Bec03} was set up at ISOLDE/CERN to measure the recoil ions after $\beta$ decay and to determine the $\beta$-$\nu$ angular correlation from the spectral shape of their energy spectrum.



\section{Experimental set-up}

\begin{figure}
\begin{center}
  \includegraphics[width=2.75 in]{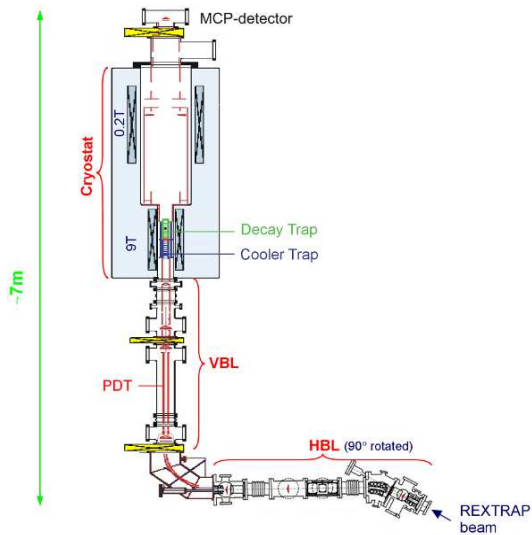}
  \caption{Schematic overview of the WITCH set-up and its
    environment. Radioactive ions received from ISOLDE are cooled and bunched
    in REXTRAP, sent into the WITCH horizontal beamline (HBL), decelerated
    in the vertical beamline (VBL) and injected into a first Penning
    trap. After cooling and, if necessary, mass selective purification of the ions in this cooler trap they are transferred to a second Penning trap where the stored ion cloud serves as scattering free source for the experiment. Both Penning traps are placed within a $9~\mathrm{T}$ solenoid. The magnetic field decreases smoothly towards the analysis plane at $0.1~\mathrm{T}$ to adiabatically collimate the recoil momentum parallel to the magnetic field lines with the magnetic gradient force. In the analysis plane this axial energy of the recoil ions is probed by a retardation potential. Recoil ions with sufficient axial energy to pass the retarding potential are accelerated with up to $-10~\mathrm{kV}$ and focussed with an Einzel lens onto a position sensitive MCP detector where they are counted. For more details see ref.~\cite{Bec03}.}
\label{fig:setup_overview}
\end{center}
\end{figure}

Most of the recent $\beta$-$\nu$ correlation experiments observe the $\beta$ particle and the recoil nucleus in coincidence (see {\it e.g.} \cite{Sci04,Gor05,Vet08,Fle08}). At the WITCH experiment, the $\beta$-$\nu$ angular correlation coefficient $a$ will be derived from the shape of the recoil energy spectrum alone. This can be done with high statistics using different isotopes independent of their chemical properties\footnote{A limitation in the choice of the isotope is the complexity of its decay and therefore the ease with which the recoil energy spectrum can be interpreted.} \cite{Bec03}. Thus, systematic effects and potential experimental artefacts can be studied in detail. The initial goal is to reach a sensitivity of $\Delta a < 0.5\%$, comparable with the best individual existing experiments. However, as the typical nuclear recoil energies after $\beta$ decay are only of the order of 100 eV,  inelastic scattering of the recoiling particles in the source is of concern. In order to avoid the latter the radioactive ions are stored in a Penning trap \cite{Bla06}. For the measurement of the recoil energy spectrum WITCH uses an electromagnetic retardation spectrometer with magnetic adiabatic collimation \cite{picard-nimb,lobashev85}. An overview of the set-up is shown in fig. \ref{fig:setup_overview}.
The radioactive ions obtained from ISOLDE, CERN \cite{Kug92} are first bunched and cooled in the REXTRAP cooler and buncher Penning trap \cite{Ame05}. They are passed on to the WITCH experiment where they are first decelerated from an energy of 30 keV to several 100 eV with the help of a pulsed drift tube (PDT, \cite{Coe07a}) before they are injected into the first of two Penning ion traps in a magnetic field of $9~\mathrm{T}$, the cooler trap of WITCH. In this trap the ion cloud is cooled by helium buffer gas and mass selectively purified \cite{Sav91,Coe07b}. The ions are then transferred to the second Penning trap, the decay trap, in which they are held either until they decay or until the trap is emptied for a subsequent measurement cycle. The decay trap is separated from the cooler trap by a differential pumping barrier to provide a scattering free source. It was operated at a depth of $\approx 8~\mathrm{V}$ for the measurements presented in the following. The recoil ions are emitted isotropically in the decay trap and in general have a maximum kinetic energy of $\mathcal{O}(100~\mathrm{eV})$. Those recoil ions from $\beta$ decay with an axial energy that is larger than the depth of the decay trap will leave the trap and move from the high magnetic field of $9~\mathrm{T}$ in the decay trap into the weak magnetic field of $0.1~\mathrm{T}$ at the analysis plane where an electric retarding potential $U_{ret}$ is applied. Due to the magnetic gradient force most of the energy of the particles is in the longitudinal component once they reach the weak field region and thus can be probed there by the retardation potential. This is the principle of a MAC-E filter \cite{picard-nimb,lobashev85}, which is used at other experiments to determine the neutrino mass by observing the $\beta$ decay of tritium \cite{FZK04,Bec10}. All ions that cross the retardation barrier are detected with a 47 mm diameter position sensitive microchannel plate detector \cite{Lie05,Coe06}. To achieve a good detection efficiency the ions are accelerated with up to $-10~\mathrm{kV}$ between the analysis plane and the detector. An Einzel lens focusses the accelerated ions onto the detector. By varying the retardation potential $U_{ret}$ an integral energy spectrum of the recoil ions is obtained.

\section{Test case: $^{124}\mathrm{In}$}

In the past few years the WITCH set-up has been extensively tested and optimized
\cite{Koz06,Koz09}. In order to test the operation of WITCH and to measure a first recoil energy spectrum an isotope which decays via beta-minus decay and which has a low ionization potential, $^{124}\mathrm{In}$, was chosen. Indium has a very low ionization potential ($\Phi_{\mathrm{In}}= 5.8~\mathrm{eV}$), which prevents losses from the decay trap caused by charge exchange with rest gas ($\Phi_{\mathrm{He}}= 24.6~\mathrm{eV}$, $\Phi_{\mathrm{N}}= 14.5~\mathrm{eV}$, $\Phi_{\mathrm{O}}= 13.6~\mathrm{eV}$). The Q-value of the $\beta^-$ decay of $^{124}In$ is $7360~\mathrm{keV}$. The highest recoil energy of $E_{rec}=196~\mathrm{eV}$ stems from decays to the first excited state at $1131.6~\mathrm{keV}$, since no decay occurs to the ground state. This excited state decays within less than $1~\mathrm{ps}$ to the ground state. The recoil due to the emitted $\gamma$ particle increases the maximum recoil energy to $E_{max}=267~\mathrm{eV}$. As a complication, $^{124}\mathrm{In}$ has an isomer, $^{124m}\mathrm{In}$, with an excitation energy of $50~\mathrm{keV}$. The $\beta$ decay with the highest Q-value also occurs to excited states only, with the lowest state at $4838~\mathrm{keV}$, resulting in a maximum recoil energy of $123~\mathrm{eV}$ and $268~\mathrm{eV}$ including the $\gamma$ recoil. The half-lives of the two Indium isomers are $t_{1/2}(^{124m}\mathrm{In})=3.7 \pm 0.2~\mathrm{s}$ and $t_{1/2}(^{124}\mathrm{In})=3.11 \pm 0.10~\mathrm{s}$. $^{124}$In is produced in large amounts at ISOLDE ($>10^8~\mathrm{ions/s}$). Thus, potentially still low efficiencies at WITCH do not pose a problem. Furthermore, the $\beta^-$ decay of the singly charged $^{124}\mathrm{In}$ ions stored in the decay trap will lead to predominantly doubly-charged positive ions and thus offers a count rate higher by an order of magnitude in comparison to beta-plus decay, for which the daughter of the decay of singly charged ions will be neutral and one has to rely on shake-off to get charged recoil ions\footnote{After the $\beta^-$ decay of neutral atoms typically about 10\% of the daughter ions will be accompanied by one shake-off electron (see {\it e.g.} \cite{Car68}). For the simple estimate above it is assumed that this is also the case for the decay of singly charged ions. This will be investigated in detail in the future.}.

Both Indium isomers have a complex decay scheme and produce a high $\gamma$ ray background, which caused discharges in the combined electric and magnetic fields of the spectrometer and the acceleration section. Therefore, the acceleration and focussing electrodes had to be operated at voltages significantly below their design values, leading to some transmission losses of the ions. As another consequence the retardation spectrometer could not be used as intended due to discharges at the main spectrometer electrode. Instead, the Einzel lens behind the main spectrometer electrodes, which is normally only used to focus the ions onto the detector, was used as a temporary retardation electrode. Both changes with respect to the design parameters of WITCH were acceptable for a first measurement of recoil ions but naturally prevented a precise measurement of their energy spectrum. Off-line measurements after the Indium measurement with a $\gamma$ ray source ($^{60}\mathrm{Co}$) at the place of the ion cloud confirmed that these discharges were indeed due to a high $\gamma$ background \cite{Koz09} and therefore are expected to be absent with an isotope with lower $\gamma$ multiplicity.

For the first measurement of recoil ions the Penning traps were operated in a magnetic field of $B_{trap}=6~\mathrm{T}$ with a depth of $U_{trap} \approx 8V$. The magnet for the definition of the B-field in the analysis plane was set to $B=0.1~\mathrm{T}$, resulting in a field in the center of the Einzel lens that was used as temporary retardation electrode of $B_{Einzel} \approx 0.01~\mathrm{T}$.

\section{Results}

\subsection{Recoil ions from $\beta$ decay in the decay trap}

\begin{figure}
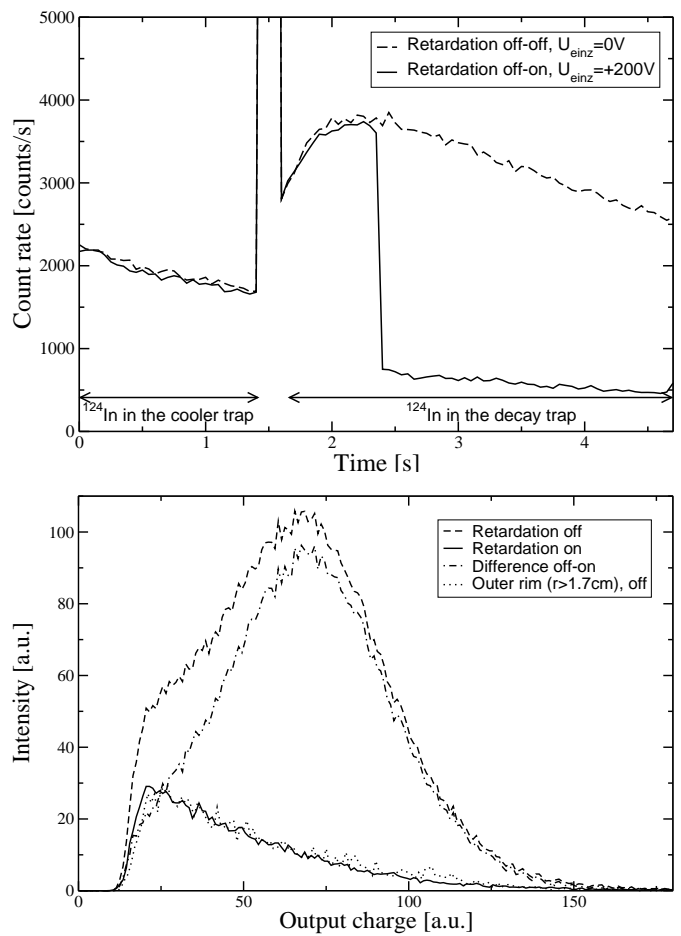

  \includegraphics[width=0.485\textwidth, angle = 0]{fig2a.eps}
\vskip 0.2cm
  \includegraphics[width=0.485\textwidth, angle = 0]{fig2b.eps}
  \caption{Top: Count rate throughout the measurement cycle during an \textit{off-on} measurement. For the case \textit{off-on} the retardation was switched \textit{on} at time $t=2.4~\mathrm{s}$, leading to a clear drop in count rate as the recoil ions are being reflected by the retardation potential (see text for details). Bottom: Pulse height distributions (PHD) obtained from the MCP detector
  during an \textit{off-on} measurement. An exponentially shaped pulse height
  distribution is characteristic for $\beta$ particles, a bell shaped distribution for ions. For retardation \textit{off} both components are visible, for
  retardation \textit{on} just the background contribution, which has the
  shape expected for $\beta$ particles. The difference of the two exhibits the
  shape expected for ions. Comparing the PHD for retardation \textit{on} with
  that on the outer rim of the detector for retardation \textit{off} clearly
  shows that the background is consistent with $\beta$ particles and that the
  ions are focussed towards the center of the detector and are absent at higher
  radii (see also fig.~\ref{fig:exppos}).
}\label{fig:on-off}
\end{figure}

The first measurement performed was an \textit{on-off} measurement for which the retardation voltage was switched between two extreme values: $U_{ret}=0~\mathrm{V}$ (\textit{off}), letting all recoil ions pass, and $U_{ret}=200~\mathrm{V}$ (\textit{on}), by far large enough to retard all recoil ions from the $\beta^-$ decay of singly charged $^{124}\mathrm{In}$, which are at least doubly charged. A significant decrease of the count rate at the moment of the switching from \textit{off} to \textit{on} then indicates the presence of positive particles of low energy, {\it i.e.} ions. The measured count rate throughout this measurement cycle is shown at the top of fig.~\ref{fig:on-off}.
The cycle starts when the ions are injected into the cooler trap. After 1.5 s they are transferred to the decay trap. Note that some ions were not captured in the decay trap and ended up on the detector, explaining the momentary increase in count rate at 1.5 s. In the \textit{off-on} mode the retardation was switched on after 2.4 s. The curve marked \textit{off-off} in fig.~\ref{fig:on-off} shows the count rate for the mode where the retardation potential was not switched but otherwise used the same measurement cycle.  As expected, switching the retardation voltage $U_{ret}$ {\it on} results in a much lower count rate. The difference is attributed to ions with $E/q < U_{ret} = 200~\mathrm{V}$ and a signal to noise ratio of about 4:1. The \textit{off-off} measurement was derived from 60 trap loads and the \textit{off-on} measurement from 90 trap loads. Each had 95 time steps with 50ms per step. The ions that hit the detector at 1.5 s caused a significant dead time of the detector during almost a second as can be seen from the variation of the count rate between 1.6 s and 2.4 s after the start of the measurement cycle in fig.~\ref{fig:on-off} (top); see also \cite{Coe06} for more details. 

The pulse heights obtained from the detector during both measurement periods are shown at the bottom of fig.~\ref{fig:on-off}. The exponentially decreasing pulse height distribution (PHD) observed during the retardation \textit{on} period ($U_{ret}=200~\mathrm{V}$) is consistent with the PHD caused by $\beta$ particles as observed in off-line tests. The additional bell-shaped distribution during the retardation \textit{off} period ($U_{ret}=0~\mathrm{V}$) is consistent with the PHD of ions on microchannel plates \cite{Lie05}.

\begin{figure}[htb]
\begin{center}
  \includegraphics[height=3 in, angle = 270]{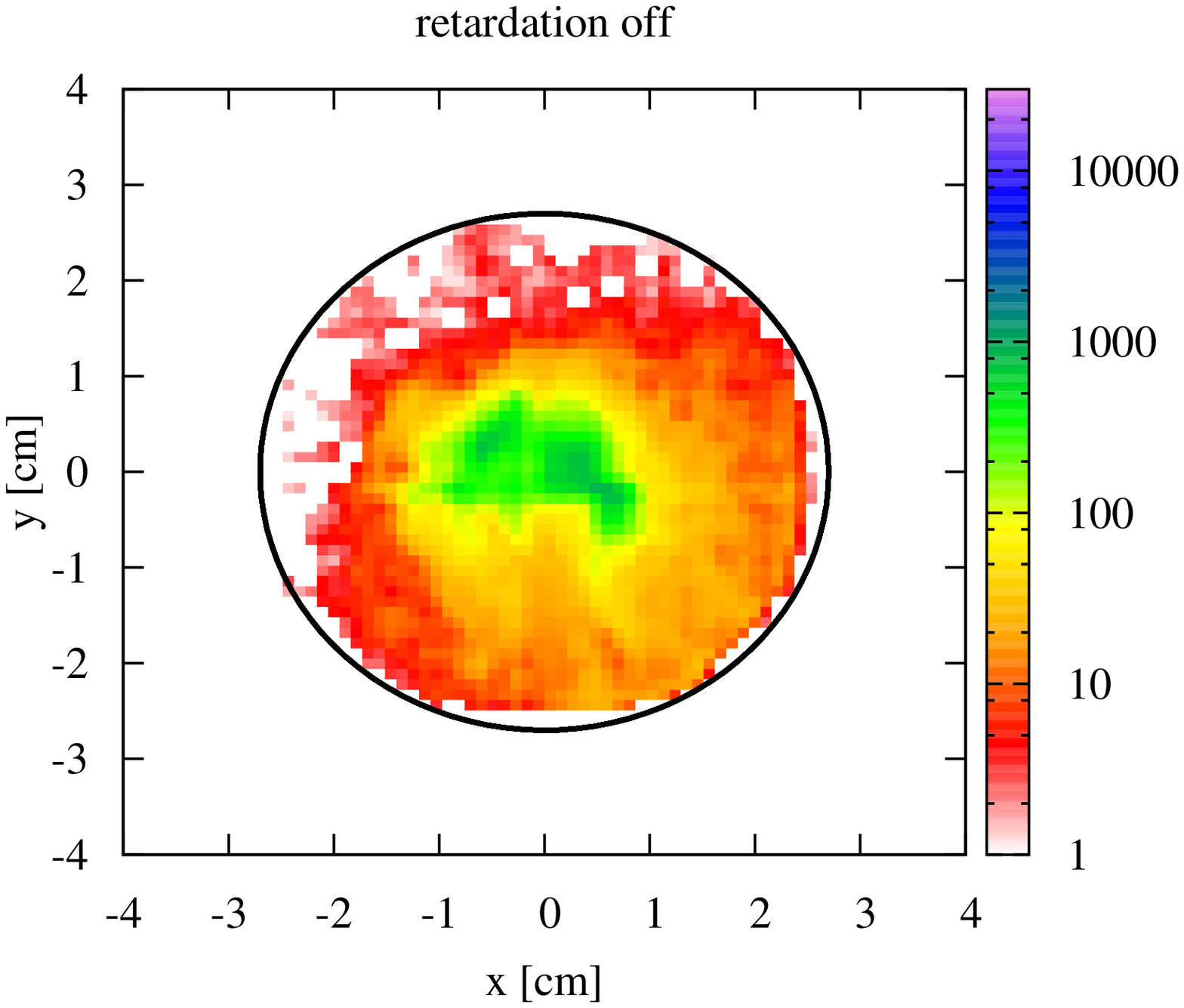}
  \includegraphics[height=3 in, angle = 270]{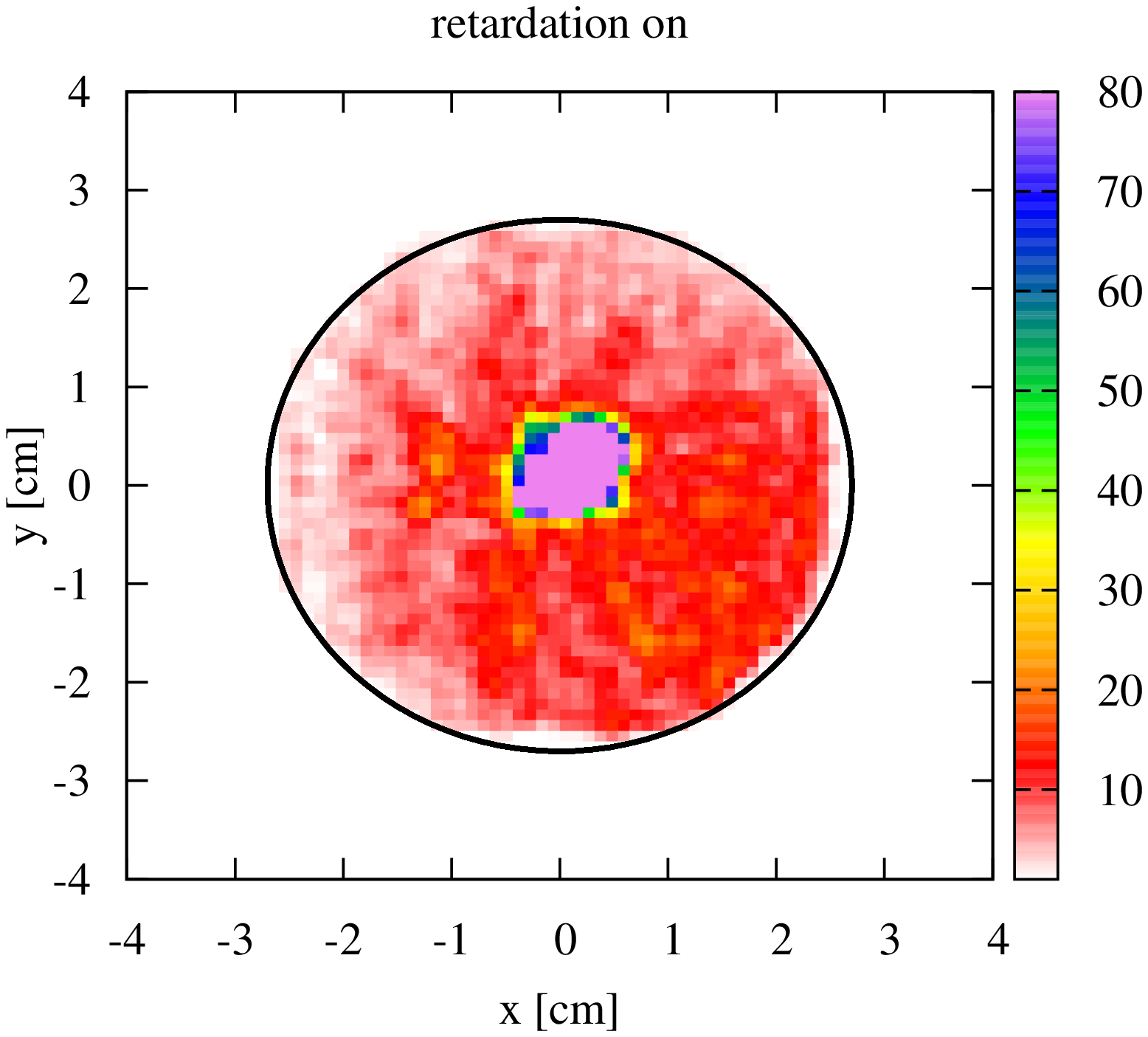}
  \caption{Measured position distribution on the detector for retardation \textit{off} ($U_{ret}=0~\mathrm{V}$, top) and retardation \textit{on} ($U_{ret}=200~\mathrm{V}$, bottom). The \textit{off} distribution shows both ions and $\beta$ particles, the \textit{on} distribution $\beta$ particles only, execpt for the center where decays of ions were measured that were deposited on the detector during the loading of the decay trap due to non-optimal trapping efficiencies. The black circle denotes the extent of the detector with diameter 47 mm. Figure~\ref{fig:simpos} shows the simulated position distribution for comparison.}
  \label{fig:exppos}
\end{center}
\end{figure}

\begin{figure}[hbt]
\begin{center}
  \includegraphics[height=3 in, angle = 270]{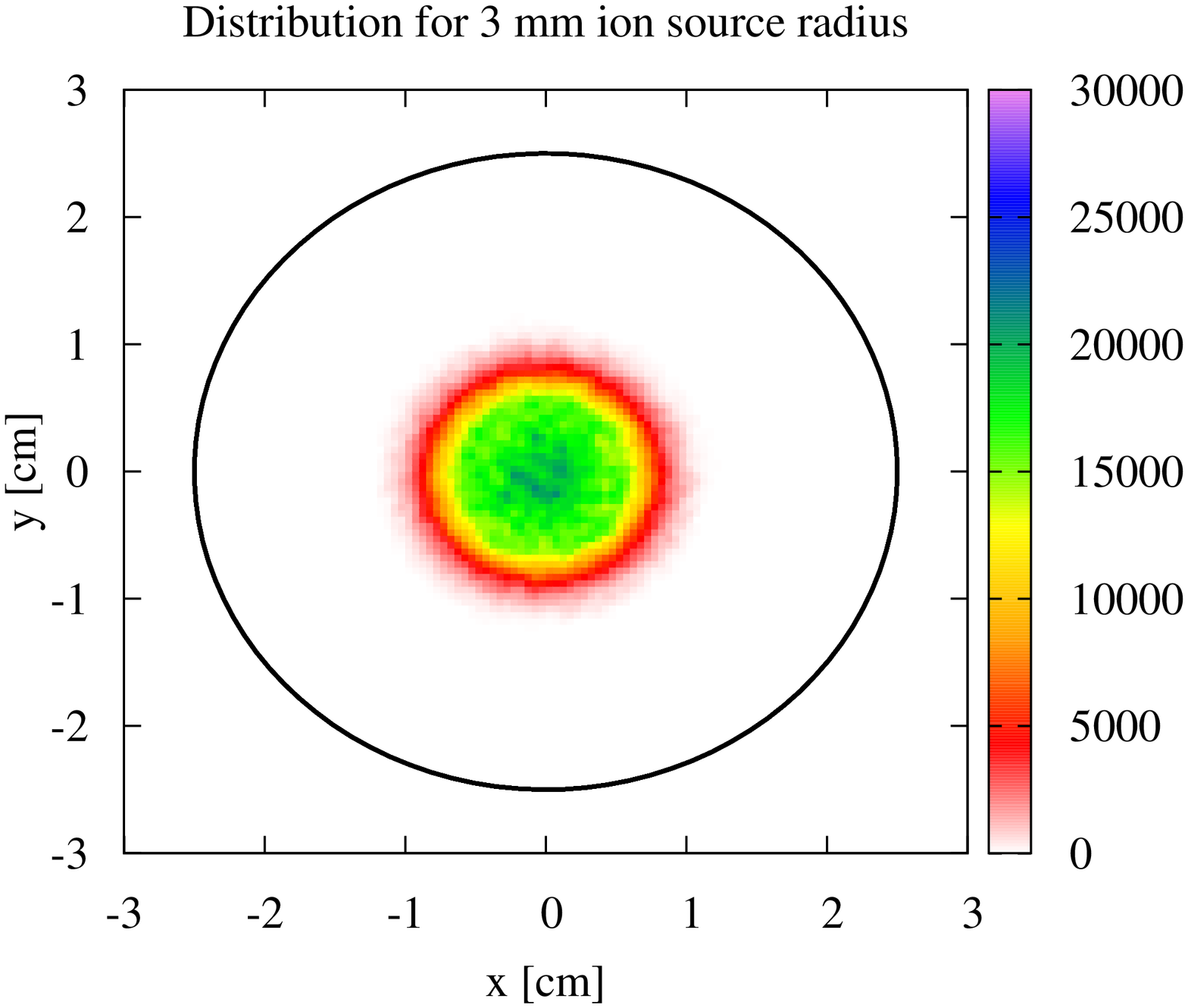}
  \includegraphics[height=3 in, angle = 270]{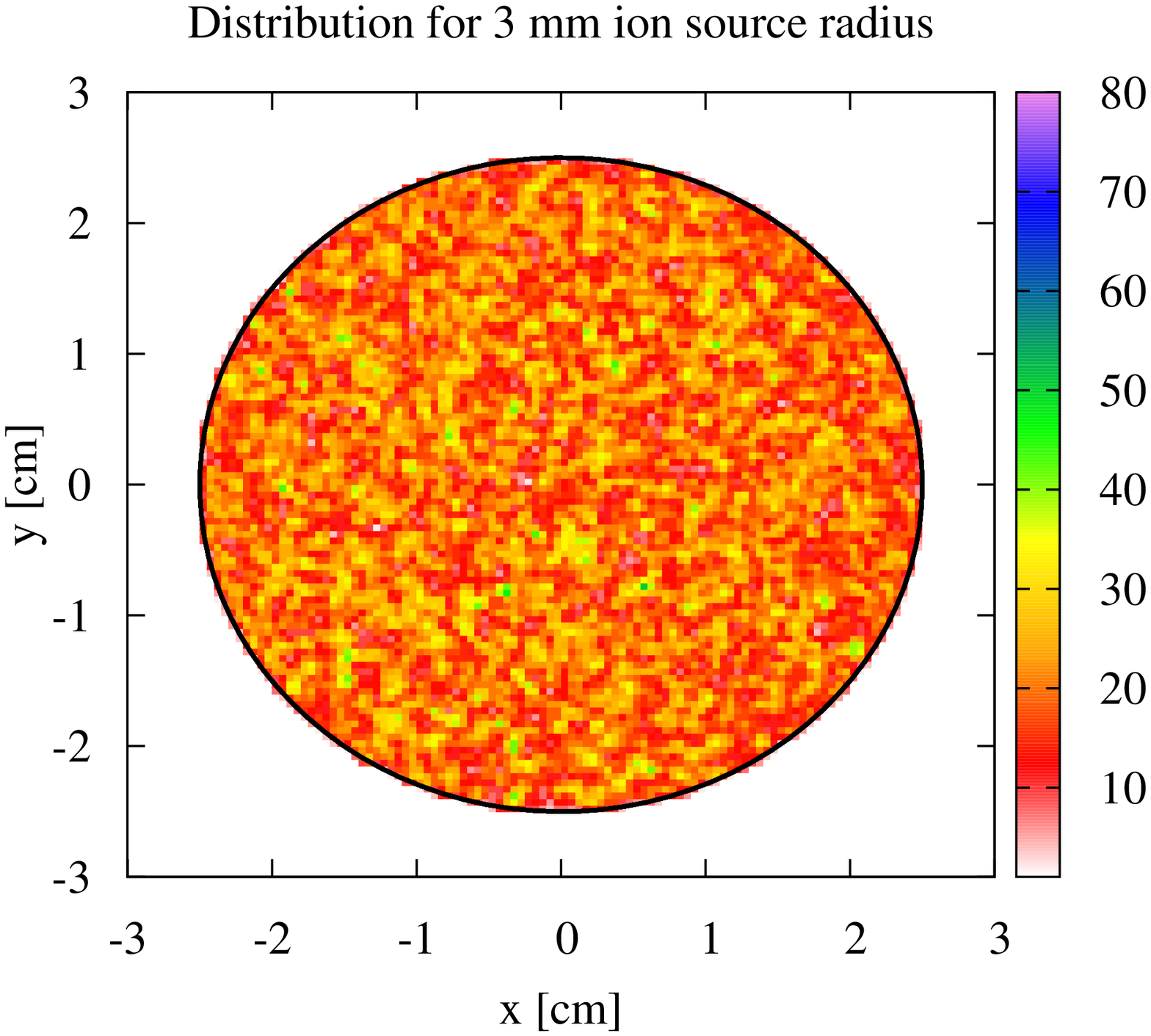}
  \caption{Simulated position distribution of recoil ions (top) and $\beta$ particles (bottom) from the decay trap on the detector. In contrast to fig.~\ref{fig:exppos} (top) the simulated distribution of the recoil ions does not include $\beta$ particles. The black circle denotes the extent of the detector. To arrive at these distributions microscopic tracking simulations of recoil ions and $\beta$ particles from a cylindrical ion cloud with constant density were performed from the decay trap to the detector. The electrode potentials were those used during the measurements. The simulations are for an ion cloud radius of $3~\mathrm{mm}$, for ions of charge state $q=2^+$ and spectra corresponding to $a=-1/3$ using the maximum endpoint energies.}
  \label{fig:simpos}
\end{center}
\end{figure}

Besides the PHDs also the spatial distribution of the events on the detector was measured. It shows a pronounced peak in the center of the detector when both the ion and the $\beta$ particle component are visible during the \textit{off} measurement ($U_{ret}=0~\mathrm{V}$, fig.~\ref{fig:exppos}, top). In contrast, only a flat component with a narrow peak remains when the retardation is switched \textit{on} ($U_{ret}=200~\mathrm{V}$) and no ions from the $\beta$ decay of the $^{124}\mathrm{In}$ in the decay trap will reach the detector (fig.~\ref{fig:exppos}, bottom). The pronounced wide peak with retardation \textit{off} can be described by the focussing of recoil ions from the trap onto the detector (fig.~\ref{fig:simpos}, top). The narrow peak for retardation \textit{on} is due to radioactive ions that were shot onto the detector upon transfer and their decay on the MCP surface. The flat backgound across the full detector can in both cases be attributed to $\beta$ particles from decays in the decay trap (fig.~\ref{fig:simpos}, bottom). This is confirmed by the pulse height distribution of the events in the outer rim ($r>1.7$ cm) of the detector in fig.~\ref{fig:on-off} (bottom panel), which has exactly the same exponential shape as the pulse height distribution for the events observed with the retardation \textit{on} ($U_{ret}=200~\mathrm{V}$) when only $\beta$ particles can reach the detector.

A fit of the half-life (see table~\ref{tab:halflife}) of the exponential decrease of the count rate when the radioactive ions are inside the decay trap and the retardation is \textit{off}, shows that the half-life is consistent with that of a mixture of $^{124}\mathrm{In}$ and $^{124m}\mathrm{In}$ and that, except for the radioactive decay, no significant loss of ions occurs during the 2.4 s measurement period.

In summary, the difference of the observed count rates for the retardation potential \textit{on} and \textit{off} shows that low energy positive particles are detected. Their PHD shows that these particles are ions. The position distribution, especially in conjunction with the tracking simulations, further shows that these ions are consistent with ions from the decay trap. Together with the analysis of the half-life this confirms that these are recoil ions from the $\beta$ decay of $^{124g,m}\mathrm{In}$ in the decay trap, which are accompanied by a diffuse background of $\beta$ particles.

\subsection{First recoil ion energy spectrum with WITCH}

\begin{figure}
\begin{center}
  \includegraphics[width=0.485\textwidth, angle = 0]{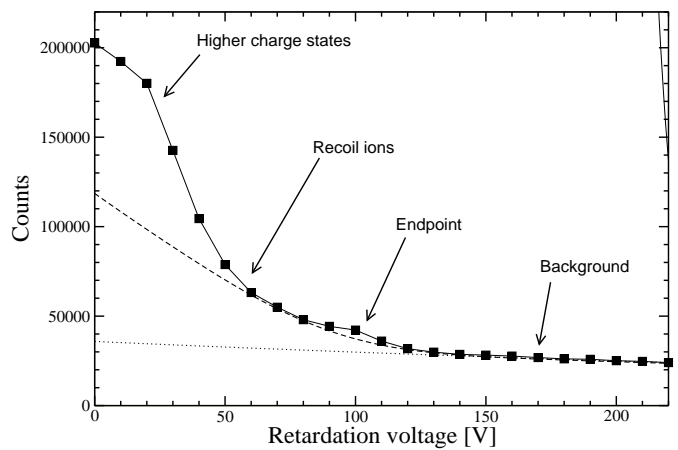}
  \caption{Measured integral energy spectrum for the recoiling ions from the $\beta$ decay of $^{124g,m}\mathrm{In}$. The retardation voltage was scanned from $0~\mathrm{V}$ to $220~\mathrm{V}$ in steps of $10~\mathrm{V}$ for this first measurement of a recoil spectrum at WITCH. At each step the number of events is measured for $100~\mathrm{ms}$. The spectrum consists of the sum of 500 trap loads, {\it i.e.} each data point has an effective measurement time of $50~\mathrm{s}$. The raw data are shown, {\it i.e.} the exponential decay has not been unfolded (squares). The data points are connected by a line to guide the eye.
For comparison a calculated recoil energy spectrum is shown for the lowest expected charge state $q_{min}=+2$ and for the maximal Q-value possible (dashed line). The calculation includes the exponential decay as well as a Gaussian broadening to approximate the thermal Doppler broadening due to high kinetic energies of the decaying parent in the trap and the Doppler broadening due to $\gamma$ decays of the excited daughter state in flight.
The constant background, which in this representation decreases with the decay constant of the $^{124}\mathrm{In}$-decay, is also shown (dotted line). At low retardation voltage the data overshoot this simple calculation significantly. This may be explained by ions of higher charge states $q$, which are created due to electron shake-off, and which increasingly contribute at lower retardation voltages according to $U_{max}(q) = E_{max}/q$. Close to the theoretical endpoint at $E_{max}=196~\mathrm{eV}$, {\it i.e.} $U_{max} = E_{max}/q_{min} = 98~\mathrm{V}$, the data overshoot the calculation somewhat, showing that a more complex model is needed to correctly describe the data. The effects leading to deviations from the calculations still have to be understood in detail and have to be minimized for future measurements.}
  \label{fig:124In_spectrum_fit}
\end{center}
\end{figure}

In order to obtain a first recoil energy spectrum with WITCH the retardation potential was scanned in 23 steps during each trap load with a step size of $10~\mathrm{V}$ and a step duration of $100~\mathrm{ms}$ (fig.~\ref{fig:124In_spectrum_fit}). Provided that there are no losses of ions from the decay trap this measurement cycle, for which all retardation voltages are scanned during one trap load, starting at $t=0~\mathrm{s}$ with $U_{ret}=0~\mathrm{V}$, implicitely takes care of the normalization. As a consequence, in the analysis any such measured spectrum has to be corrected for the half life due to the decay during the measurement. Such a measurement cycle also implies that the abscissa in fig.~\ref{fig:124In_spectrum_fit} corresponds to time as well as retardation potential.

Since the singly charged $^{124}\mathrm{In}$ and $^{124m}\mathrm{In}$ ions decay via $\beta^-$ decay the recoil ions are at least doubly charged ($q_{min}=2$). Therefore, without any additional energy broadening, a retardation potential of $U_{ret}=E_{max}/q_{min} = 98~\mathrm{V}$ is sufficient to retard all recoil ions and constitutes the theoretical endpoint of the recoil energy spectrum (as indicated in fig.~\ref{fig:124In_spectrum_fit}). In this case all events measured above this voltage are background due to $\beta$ particles from the $\beta$ decay. In practice, both Doppler broadening due to $\gamma$ decay in flight and thermal Doppler broadening due to a non-vanishing kinetic energy of the radioactive ions in the decay trap will result in some events above the theoretical endpoint energy, as is visible in fig.~\ref{fig:124In_spectrum_fit}. The $\gamma$ ray broadening will lead to an endpoint of $E_{max,\gamma} = 267~\mathrm{eV}$ and $U_{ret,\gamma} = E_{max,\gamma}/q_{min} = 134~\mathrm{V}$. The distribution of the kinetic energy in the decay trap is unknown for this Indium measurement, but the average kinetic energy in the trap can be estimated to $\mathcal{O}(\mathrm{eV})$ based on the trap potential necessary to trap all ions. This will result in a thermal Doppler broadening of $\mathcal{O}(20~\mathrm{eV})$, \textit{i.e.} an additional increase of the retardation voltage up to which recoil ions can be detected of $\mathcal{O}(10~\mathrm{V})$. In addition to the minimal charge state $q_{min}=2$ also higher charge states $q$ may be present, when taking electron shake-off into account. They contribute to the spectrum below $U_{ret}=E_{max}/q$ accordingly.

An in-depth analysis of the measured recoil spectrum is not possible due to several experimental artifacts: i.) non-ideal acceleration and focussing potentials because of discharges in the acceleration section and at the detector caused transmission losses. ii.) the ions which were shot directly onto the detector caused a significant dead time of the detector, different for the different times in the measurement cycle. iii.) for this first measurement of a recoil energy spectrum just a few retardation steps were taken. They are not sufficient for fitting all free parameters.

After the measurements with $^{124}\mathrm{In}$ additional measurements were performed with a $\beta$ and a $\gamma$ source each in the center of the decay trap (viz. $^{90}\mathrm{Sr}$ and $^{60}\mathrm{Co}$). These resulted in mostly flat spectra \cite{Koz09}, {\it i.e.} totally different spectral shapes than the one shown in fig.~\ref{fig:124In_spectrum_fit}, confirming that the ions measured with $^{124}\mathrm{In}$ in the decay trap were not caused by spurious effects but stem from $\beta$ decays in the decay trap \cite{Koz09}.

\subsection{Properties of the ion cloud in the decay trap}

Besides the confirmation of the observation of recoil ions from the decay trap several experimental parameters which influence the performance and the achievable sensitivity of WITCH can be deduced from the above measurements. These are i.) the half-lives of the decaying isotopes and their storage time in the decay trap, which yield information about the proper operation of the trap, ii.) the number of ions in the ion cloud, which, together with the signal-to-noise ratio, is important for the statistics of the measurement and therefore the achievable sensitivity of WITCH, iii.) the size of the ion cloud, which influences the size of the image of the recoil ions on the detector and iv.) the alignment of the whole set-up from the traps to the detector.

\subsubsection{Half-life and losses from the ion cloud}
\label{sec:halflife}

The count rate during the measurement of one trap load with constant retardation potential should show a decrease corresponding to the half-life of the isotope investigated. In the present case a mixture of $^{124}\mathrm{In}$ and $^{124m}\mathrm{In}$ is measured and the observed half-life is determined by their relative abundance in the ion cloud. An additional decrease of the count rate beyond what is expected from the half-life of the radioactive decay would be a sign of a loss of ions from the decay trap and would lead to significant systematic uncertainties\footnote{For example any loss of ions from the trap would lead to a lower efficiency for high recoil energies than for low recoil energies due to the measurement cycle used. This would induce a change of the shape of the recoil energy spectrum and thus of the $\beta$-$\nu$ angular distribution that is extracted from it.} for the recoil energy spectrum for the measurement cycle discussed above.

Table~\ref{tab:halflife} shows the experimentally determined decay constants for the \textit{off-on} and \textit{off-off} measurements shown in fig.~\ref{fig:on-off} for the signals coming from the cooler trap in the first 1.4 seconds as well as from the decay trap in the last 1.65 seconds of the measurement cycle.

\begin{table}[htb]
\centering
\caption{Fits of the exponential decay $N(t) = N_o e^{-\lambda t}$ for the \textit{off-on} and \textit{off-off} measurements of fig.~\ref{fig:on-off} (top panel). In the time interval [0s,1.4s] the events stem from decays in the cooler trap, in [1.5,4.7] from decays in the decay trap. Events from the decay trap were only evaluated in the interval [3s,4.7s], when the MCP had recovered from the dead time caused by the ions that were shot onto it during the transfer to the decay trap. The measurement {\it off-on} in [3s,4.7s] consists of background only ($\beta$ particles according to PHD and position distribution), since all recoil ions will be reflected by the retardation potential. The measurement {\it off-off} in [3s,4.7s] consists of both signal (ions) and background ($\beta$ particles). All four regions should show the decay constant of a mixture of the two Indium isomers. The uncertainties shown are statistical only.}
\label{tab:halflife}
\begin{tabular}{|l|l|c|c|c|}
\hline
Meas. & Region & $N_o [1/50ms]$ & $\lambda$ [$\mathrm{s}^{-1}$] & $\chi^2/\nu$ \\
\hline
{\it off-off} & [3s,4.7s[ & $18020 \pm 260$ & $0.182 \pm 0.004$ & 0.98\\
{\it off-on} & [3s,4.7s[ & $5340\pm 230$ & $0.205 \pm 0.012$ & 2.5\\
{\it off-off} & [0s,1.4s] & $6635 \pm 37$ & $0.195 \pm 0.007$ & 1.5\\
{\it off-on} & [0s,1.4s] & $9749 \pm 71$ & $0.201 \pm 0.010$ & 3.7\\
\hline
\end{tabular}
\end{table}

The four measured decay constants are consistent with each other and result in a weighed average of $\bar{\lambda} = 0.188 \pm 0.003~\mathrm{s^{-1}}$, corresponding to a half-life of $t_{1/2} = 3.69 \pm 0.06~\mathrm{s}$ and an abundance of $^{124g}\mathrm{In}$ ($\lambda= 0.223 \pm 0.008~\mathrm{s^{-1}}$) to  $^{124m}\mathrm{In}$ ($\lambda= 0.187 \pm 0.011~\mathrm{s^{-1}}$) of $0.03 \pm 0.33$. Note that due to the large uncertainty of the literature values of the half-lives this cannot be determined with high precision.

The experimental decay constant of $\lambda_{ion}=0.182\pm0.004~\mathrm{s^{-1}}$ for the case when ions dominate the count rate (first line in table~\ref{tab:halflife}) is consistent with the shortest possible decay constant from $^{124}\mathrm{In}$, {\it i.e.} $\lambda(^{124m}\mathrm{In})=  0.187 \pm 0.011~\mathrm{s^{-1}}$, consistent with zero losses from the trap. A conservative upper limit on the losses from the decay trap can then be calculated to be $1/N_o~dN_{DT}/dt < 0.026~\mathrm{ s^{-1}} (CL=99.5\%)$, {\it i.e.} less than $2.6\%$ per second\footnote{At a later time a test measurement with Argon ($\Phi_{Ar}=15.8~\mathrm{eV}$) was performed. As with the Indium the Argon was cooled by Helium buffer gas in the cooler trap. In contrast to the measurements with Indium the Argon showed a rapid charge exchange in the cooler trap with an overall half-life of $t_{1/2}^{tot}(^{35}\mathrm{Ar}) \approx 8~\mathrm{ms}$ while the half-life for the $\beta$ decay is $t_{1/2}(^{35}\mathrm{Ar}) \approx 1.78~\mathrm{s}$. This shows that improvements of the rest gas levels in the traps are still needed for the measurement of elements with high ionization potential.}.

\subsubsection{Number of ions in the ion cloud}

A high source strength, {\it i.e.} a high number $N_{DT}$ of radioactive ions that are stored in the decay trap, is important to reach low statistical uncertainties. From the normalization constants $N_o(on)$ and $N_o(off)$, which give the event numbers in an interval of $0.05~\mathrm{s}$, follows the initial count rate in the MCP, $\frac{dN_{MCP}}{dt} = 3650 \pm 80~\mathrm{s^{-1}}$ (at $t=1.5~\mathrm{s}$, the time of the filling of the decay trap), per trap load. This count rate is related to the initial number of ions in the decay trap $N_{DT}$ as

\begin{equation}
\label{nd}
\frac{dN_{MCP}}{dt} = \epsilon_\Omega \cdot \epsilon_{Trans} \cdot \epsilon_{MCP} \cdot \lambda_{ion} \cdot N_{DT}
\end{equation}

\noindent with $\lambda_{ion}=0.182 \pm 0.004~\mathrm{s}^{-1}$ the decay constant measured for the ions, $\epsilon_\Omega = 0.5$ for the solid angle of the decays emitted into the forward direction, $\epsilon_{Trans} = 0.56 \pm 0.05$ the transmission probability for doubly charged recoil ions from the decay trap (depth of $\approx 8V$) to the detector, estimated from simulations for an ion cloud radius of $2~\mathrm{mm}$ (see sect.~\ref{sec:size}), and $\epsilon_{MCP} = 0.523$ measured for a comparable MCP \cite{Lie05} and close to the open area ratio of $0.55$, i.e. the ratio of the area of the MCP covered by channels to the area in between the channels, which defines the sensitive area of an MCP. The probability for the creation of the different charge states does not need to be included since the measurement was done at zero retardation voltage and all charge states could in principle reach the detector.

Using these numbers results in an average trap load of $N_{DT} = (1.4 \pm 0.2) \cdot 10^5$ Indium ions\footnote{Higher charge states will have a lower escape probability from the decay trap ({\it i.e}\ a lower $\epsilon_{Trans}$). However, since they contribute only at the $10-20\%$ level, this has been neglected for this estmate.}. This is a reasonable source strength for test measurements but is an order of magnitude smaller than the design value of $10^6$ ions per trap load. Thus, several features of the set-up still have to be optimized for a precision measurement, as {\it e.g.} the injection efficiency into the cooler trap, which was measured to be just $\approx 20\%$ for this initial experiment.

\subsubsection{Size of the ion cloud}
\label{sec:size}

\begin{figure}
\begin{center}
  \includegraphics[width=0.3\textwidth, angle = -90]{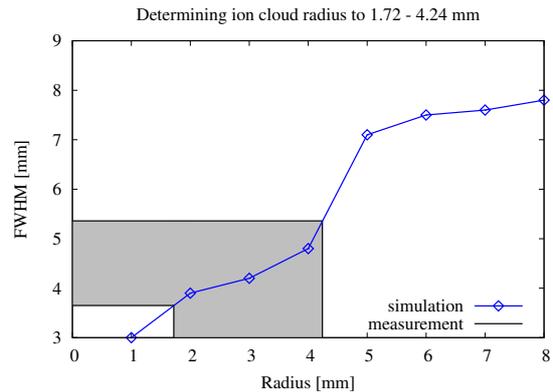}
  \caption{FWHM of the simulated position distribution of ions on the detector as a function of the size of the ion cloud in the decay trap. The shaded area marks the FWHM of the experimental position distribution. Comparing both  yields a radius of the ion cloud in the range $[1.5~\mathrm{mm},4.3~\mathrm{mm}]$. The simulations are for ions of charge state $q=2^+$ and use a recoil spectrum with the maximum endpoint energy only.}
  \label{fig:FWHMcloudsize}
\end{center}
\end{figure}

The size of the ion cloud in the decay trap is one of the factors that determine whether all recoil ions are focussed onto the detector and what their spot diameter on the detector is. The decay trap was operated in box trap mode, {\it i.e.} the endcap electrodes were at the trapping potential ($\approx 8~\mathrm{V}$) and the inner electrodes all at the same low potential ($0~\mathrm{V}$), leading to a longitudinal ion cloud size of $\approx 100~\mathrm{mm}$. The radial extent of the ion cloud can be inferred from the size of the ion spot on the detector by a comparison with tracking simulations for different ion cloud sizes. Figure~\ref{fig:FWHMcloudsize} shows the comparison of the simulated spot sizes with the measured one. For the simulation a cylindrical ion cloud of homogeneous density was assumed\footnote{A detailed model for realistic density distributions at different trap potentials and number of ions still has to be developed.}. The comparison results in an estimate of the ion cloud radius in the range $[1.7~\mathrm{mm},4.3~\mathrm{mm}]$. This is larger than the radius of the differential pumping barrier of $1.5~\mathrm{mm}$, which determines the initial maximal size of the ion cloud. The simulated position distributions for the ions and the $\beta$ particles for a radius of $3~\mathrm{mm}$ are shown in fig.~\ref{fig:simpos}.

\subsubsection{Signal to noise ratio}

The ion cloud size also determines the signal to noise ratio, since in the
ideal case all recoil ions are focussed onto the detector whereas the number of
$\beta$ particles that hit the detector decreases with increasing ion cloud size
because electrons with higher starting radii have a smaller
probability of hitting the detector and will eventually not hit the detector
anymore. The signal to noise ratio can be extracted from the fit results shown in table~\ref{tab:halflife}. For the \textit{off-on} measurement the normalization constants for the signal region are $N_o(on)=5340\pm 230$ for 90 trap loads and $N_o(off)= 18020 \pm 260$ for 60 trap loads, yielding a signal to noise ratio of $S/N = 4.06 \pm 0.23$.

The background can be further subdivided into background from $\beta$ particles, which are distributed across the full detector surface, and background due to decays on the detector surface of those ions that were shot onto the detector upon transfer, which appear as a peak close to the center of the detector ($\textit{on}$ measurement, fig.~\ref{fig:exppos}, bottom panel). These two contributions can be separated by counting the events outside and inside a circle centered around the strongly peaked maximum in the position distribution during the \textit{on} measurement. Inside a circle of radius\footnote{The radius of the MCP detector is $23.6 \textit{mm}$.} $7.5 \textit{mm}$ are $15670$ and outside $15283$ events. When the $\beta$ background of $15670$ events is extrapolated to radii $< 7.5 \textit{mm}$ this results in a total number of events due to $\beta$ particles of $16605$ for the full MCP area and a number of events due to decays on the detector of $14347$. This means that $\approx 46 \%$ of the background are due to decays on the detector. Using this fraction to correct both the background and the signal count rates for the $\textit{off}$ measurement the signal to noise ratio can be $S/N \approx 5.2$, assuming that the trapping efficiency is improved and no radioactive ions get shot onto the detector.


\subsubsection{Alignment traps-detector}

As already mentioned, the narrow peak for the \textit{on} distribution in fig.~\ref{fig:exppos} (bottom) in the center of the detector corresponds to the decay of radioactive ions that were shot onto the detector due to a trapping efficiency $ < 100\%$. This peak can be used as a measure of the alignment and focussing properties of the WITCH set-up: The ions were shot through the differential pumping barrier, which is mechanically centered along the axis of the traps. The ions have dominantly longitudinal kinetic energy, travel parallel to the B-field and see a cylindersymmetric electric field, since they fly close to the symmetry axis.

The spot of the ions is located at a distance of $\Delta r = 2.2 \pm 0.1~\mathrm{mm}$ from the center of the detector. With a distance of $2.644~\mathrm{m}$ between the center of the decay trap and the detector this corresponds to an angle of $0.047^\circ \pm 0.002^\circ$.

\section{Conclusion}

The WITCH experiment has been set up at ISOLDE to measure the recoil energy spectrum after $\beta$ decay, from which the $\beta$-$\nu$ angular correlation will be determined for a search for scalar weak interaction. In a first measurement with the radioactive ions $^{124}\mathrm{In}$ and $^{124m}\mathrm{In}$ it has been shown that recoiling ions from the Penning trap can be efficiently detected, thus demonstrating the experimental principle. Especially, up to $10^5$ ions have been stored in one trap load in the decay trap with losses from the trap consistent with zero during the time necessary for a measurement. The radius of the ion cloud was estimated to be in the range $[1.5~\mathrm{mm},4.3~\mathrm{mm}]$. The recoil ions from the $\beta$ decays in the trap were transmitted through the spectrometer and focussed onto the detector consistent with simulations for the settings used during the measurement. It was shown that the events detected are recoil ions and a first recoil ion energy spectrum was measured in coarse steps and with a signal-to-noise ratio of $S/N = 4$, which can still be improved.

After having observed recoil ions at WITCH, the set-up is now being further improved and systematic effects are being studied in preparation of a precise determination of the $\beta$-$\nu$-angular correlation coefficient $a$.
The experimental improvements include the elimination of discharges in the spectrometer and the reacceleration section, increasing the efficiencies of the traps, that do not capture all ions but let some get transmitted onto the detector, and the reduction of charge exchange of stored ions with high ionization potential. For this purpose a number of spectrometer electrodes have been remachined, all electrodes have been electropolished, a new set of Penning traps has been installed, all materials incompatible with the desired UHV conditions have been removed and getter vacuum pumps based on non-evaporative getter have been installed.

 For a precision analysis of the recoil spectrum the systematic problems of the set-up have to be understood as well. To this end, simulations for the traps and the spectrometer (see $e.g.$ \cite{SVG10}) as well as test measurements are ongoing. In addition, physics effects like the charge state distribution after $\beta$ decay (see {\it e.g.} \cite{Car68,Sci03}) and the decay of a potentially unstable daughter in flight will be investigated in the future.

\section{Acknowledgements}

We would like to thank the LPC-Caen low energy weak interaction group for the loan of the MCP detector.

This work is supported by the European Union grants FMRX-CT97-0144
(the EUROTRAPS TMR network) and HPRI-CT-2001-50034 (the NIPNET RTD
network), the European Union Sixth Framework through RII3-EURONS
(contract no. 506065), the Flemish Fund for Scientific Research
FWO, project GOA 2004/03 of the K.U. Leuven and by the GERMAN BMBF under Grant
No. 06MS270.


\begin{thebibliography}{99}
\bibitem{Sev06} N. Severijns, M. Beck and O. Naviliat-Cuncic, Rev. Mod. Phys. \textbf{78} (2006) 991 doi:10.1103/RevModPhys.78.991
\bibitem{Jac57} J.D. Jackson, S.B. Treiman and H.W. Wyld, Nucl. Phys. \textbf{4} (1957) 206 doi:10.1016/0029-5582(87)90019-8
\bibitem{Car91}A.S. Carnoy, J. Deutsch, T.A. Girard and R. Prieels, Phys. Rev. C \textbf{43} (1991) 2825 doi:10.1103/PhysRevC.43.2825
\bibitem{Qui93}P.A. Quin, J. Deutsch, T.E. Pickering, J.E. Schewe and P.A. Voytas, Phys. Rev. D \textbf{47} (1993) 1247 doi:10.1103/PhysRevD.47.1247
\bibitem{Ska94}M. Skalsey, Phys. Rev. C \textbf{49} (1994) R620 doi:10.1103/PhysRevC.49.R620
\bibitem{Sev00}N. Severijns \textit{et al.}, Hyp. Int. \textbf{129} (2000) 223 doi:10.1023/A:1012665917625
\bibitem{Abe08}H. Abele, Prog. Part. Nucl. Phys. \textbf{60} (2008) 1 doi:10.1016/j.ppnp.2007.05.002
\bibitem{Har09}J.C. Hardy and I.S. Towner, Phys. Rev. C \textbf{79} (2009) 055502 doi:10.1103/PhysRevC.79.055502
\bibitem{Pit09}J.R.A. Pitcairn \textit{et al.}, Phys. Rev. C \textbf{79} (2009) 015501 doi:10.1103/PhysRevC.79.015501
\bibitem{Wau09}F. Wauters \textit{et al.}, Phys. Rev. C \textbf{80} (2009) 062501(R) doi:10.1103/PhysRevC.80.062501
\bibitem{Beh09} J.A. Behr and G. Gwinner, J. Phys. G \textbf{36} (2009) 033101 doi:10.1088/0954-3899/36/3/033101
\bibitem{Joh63} C.H. Johnson F. Pleasonton and A.H. Snell, Phys. Rev. \textbf{132}(1963) 1149 doi:10.1103/PhysRev.132.1149
\bibitem{Sci04} N.D. Scielzo \textit{et al.}, Phys. Rev. Lett. \textbf{93} (2004) 102501 doi:10.1103/PhysRevLett.93.102501
\bibitem{Gor05} A. Gorelov \textit{et al.}, Phys. Rev. Lett. \textbf{94} (2005) 142501 doi:10.1103/PhysRevLett.94.142501
\bibitem{Ade99} E.G. Adelberger \textit{et al.}, Phys. Rev. Lett. \textbf{83} (1999) 1299 doi:10.1103/PhysRevLett.83.1299
\bibitem{Bec03} M. Beck \textit{et al.}, Nucl. Instr. Meth. A \textbf{503} (2003) 567 doi:10.1016/S0168-9002(03)00994-X
\bibitem{Rod06} D. Rodriguez \textit{et al.}, Nucl. Instr. Meth. A \textbf{565} (2006) 876 doi:10.1016/j.nima.2006.05.165
\bibitem{Glu05} F. Gl\"uck \textit{et al.}, Eur. Phys. J. A \textbf{23} (2005) 135 doi:10.1140/epja/i2004-10057-1
\bibitem{Vet08} P.A. Vetter, J.R. Abo-Shaeer, S.J. Freedman and R. Maruyama, \textit{et al.}, Phys. Rev. C \textbf{77} (2008) 035502 doi:10.1103/PhysRevC.77.035502.
\bibitem{Fle08} X. Fl\'echard \textit{et al.}, Phys. Rev. Lett. \textbf{101} (2008) 212504 doi:10.1103/PhysRevLett.101.212504.
\bibitem{Bas08} S. Bae\ss ler \textit{et al.}, Eur. Phys. J. A \textbf{38} (2008) 17 doi:10.1140/epja/i2008-10660-0
\bibitem{Poc09} D. Po\v{c}ani\'c \textit{et al.}, Nucl. Instr. Meth. A \textbf{611} (2009) 211 doi:10.1016/j.nima.2009.07.065.
\bibitem{Bla06}K. Blaum, Phys. Rep. \textbf{425} (2006) 1 doi:10.1016/j.physrep.2005.10.011
\bibitem{picard-nimb} A. Picard \textit{et al.}, Nucl. Instr. Meth. {\bf B\,63} (1992) 345 doi:10.1016/0168-583X(92)95119-C
\bibitem{lobashev85} V. M. Lobashev and P. E. Spivak, Nucl. Instr. Meth. {\bf A\,240} (1985) 305 doi:10.1016/0168-9002(85)90640-0
\bibitem{Kug92} E. Kugler \textit{et al.}, Nucl. Instr. Meth. B \textbf{70} (1992) 41 doi:10.1016/0168-583X(92)95907-9
\bibitem{Ame05} F. Ames \textit{et al.}, Nucl. Instr. Meth. A \textbf{538} (2005) 17 doi:10.1016/j.nima.2004.08.119
\bibitem{Coe07a} S. Coeck \textit{et al.}, Nucl. Instr. Meth. A \textbf{572} (2007) 585 doi:10.1016/j.nima.2006.11.054
\bibitem{Sav91} G. Savard \textit{et al.}, Phys. Lett. A \textbf{158} (1991) 247 doi:10.1016/0375-9601(91)91008-2
\bibitem{Coe07b} S. Coeck \textit{et al.}, Nucl. Instr. Meth. A \textbf{574} (2007) 370 doi:10.1016/j.nima.2007.02.079
\bibitem{FZK04} The KATIRN Collaboration (j. Angrik \textit{et al.}), \textit{KATRIN Design Report 2004}, FZKA Scientific Report 7090, 2005, available online at http://bibliothek.fzk.de/zb/berichte/FZKA7090.pdf
\bibitem{Bec10} M. Beck for The KATRIN Collaboration, J. Phys.: Conf. Ser. \textbf{203} (2010) 012097 doi:10.1088/1742-6596/203/1/012097
\bibitem{Lie05} E. Li\'enard  \textit{et al.}, Nucl. Instr. Meth. A \textbf{551} (2005) 375 doi:10.1016/j.nima.2005.06.069
\bibitem{Coe06} S. Coeck \textit{et al.},  Nucl. Instr. Meth. A \textbf{557} (2006) 516 doi:10.1016/j.nima.2005.11.061
\bibitem{Koz06} V.Yu. Kozlov \textit{et al.}, Int. J. Mass Spectrom. \textbf{251} (2006) 159 doi:10.1016/j.ijms.2006.01.050
\bibitem{Koz09} V. Yu. Kozlov \textit{et al.}, Nucl. Instr. Meth. B \textbf{266} (2008) 4515 doi:10.1016/j.nimb.2008.05.150
\bibitem{SVG10} S. van Gorp, submitted to Nucl. Instr. Meth. A (2010)
\bibitem{Car68} T.A. Carlson \textit{et al.}, Phys. Rev. \textbf{169} (1968) 27 doi:10.1103/PhysRev.169.27
\bibitem{Sci03} N.D. Scielzo \textit{et al.}, Phys. Rev. A \textbf{68} (2003) 022716 doi:10.1103/PhysRevA.68.022716
\end{thebibliography}
\end{document}